\documentclass[aps,twocolumn,showpacs]{revtex4}
\usepackage{dcolumn}
\usepackage{graphicx}
\usepackage{amsmath}
\usepackage{amsfonts}
\usepackage{amssymb}
\usepackage{psfrag}
\usepackage{wrapfig}
\usepackage{subfigure}
\usepackage{makeidx}
\usepackage{bm}
\usepackage{epsf}

\begin{document}

\title{Dynamics of Rogue Wave Excitation Pattern on Stripe Phase Backgrounds in a Two-component Bose-Einstein Condensate}
\author{Li-Chen Zhao$^{1,2}$}\email{zhaolichen3@nwu.edu.cn}
\author{Liming Ling$^{3}$}
\author{Jian-Wen Qi$^{1,2}$}
\author{Zhan-Ying Yang$^{1,2}$}\email{zyyang@nwu.edu.cn}
\author{Wen-Li Yang$^{2,4}$}

\address{$^{1}$School of Physics, Northwest University, 710069, Xi'an, China}
\address{$^{2}$Shaanxi Key Laboratory for Theoretical Physics Frontiers, 710069, Xi'an, China}
\address{$^{3}$School of Mathematics, South China University of Technology, 510640, Guangzhou, China}
\address{$^{4}$Institute of Modern Physics, Northwest University, 710069, Xi’an, China}

\begin{abstract}
We study rogue wave excitation pattern in a two-component Bose-Einstein condensate with pair-transition effects. The results indicate that rogue wave excitation can exist on a stripe phase background for which there are cosine and sine wave background in the two components respectively. The rogue wave peak can be much lower than the ones of scalar matter wave rogue waves, and varies with the wave period changing. Both rogue wave pattern and rogue wave number on temporal-spatial plane are much more abundant than the ones reported before. Furthermore, we prove that the rogue wave number can be $n (n+1)/2+m (m+1)/2$ (where $m,n $ are arbitrary non-negative integers), in contrast to $n(n+1)/2$ for scalar nonlinear Schr\"{o}dinger equation described systems. These results would enrich our knowledge on nonlinear excitations in multi-component Bose-Einstein condensate with transition coupling effects.

\end{abstract}
\maketitle
\section{Introduction}
Bose-Einstein condensate (BEC) admits tuneable contact interactions between atoms which makes the condensate can be used to investigate dynamics of many different nonlinear localized waves \cite{Kevrekidis,Wu,Matuszewski}, such as bright soliton, dark soliton, and rogue wave. For BEC with attractive interactions, it admits bright soliton on zero background and rogue wave (RW) on plane wave background \cite{Bludov1,Mrwb}. For BEC with repulsive interactions, it admits dark soliton on a plane wave background \cite{Wu}. Since there is no modulational instability for repulsive case, therefore rogue wave do not exist for repulsive scalar BEC. These characters just hold for scalar BEC systems. When the internal spin degree of freedom of the particles is taken into account,
the situation becomes even more interesting and the studies were naturally extended to vector BEC or spinor BEC systems \cite{spinbec,Becker,Engels}. It has been shown that there are many different localized waves for vector systems, in contrast to the scalar ones  \cite{Guo,Baronio,Zhao,Zhao1,Zhao2,Lakshmanan,Bludov}. This comes from that vector BEC admit the intra and external interactions between atoms which are much abundant than scalar systems.
Generally, the population or particle number in each component is conserved in the most previous studies. However, in
practical physical systems, the particle numbers in each component
are not necessarily conserved. For instance, in microscopic particle
transport, the particle in one component can transfer to another component through quantum tunneling or coherent coupling effects \cite{Li,Williams,Fischer,Liu,Wadati,Qin,Lahini}.

There are usually two transition paths for atoms:  single particle and pair particles transition \cite{Fischer1,Fischer2}. For weak contact interaction strength cases, single particle transition play dominant role for particle transition and pair particles transition can be ignored. But the pair transition effects become dominant for strong interaction strength cases \cite{pair,Meyer}. Therefore, we
consider that the case for second-order transition is dominant, for which the dynamical equation can be written as an integrable coupled nonlinear Schr\"odinger equation with pair transition coupling effects (CNLS-p) \cite{zhao-ling}. Some types of nonlinear excitations for similar integrable model have been obtained based on special Hirota bilinearization and the Darboux transformation (DT) \cite{Park}, such as bright soliton \cite{Tian}, RW \cite{BoTian}. Furthermore, we presented two Darboux transformation forms for deriving nonlinear localized wave solutions \cite{lingzhao}. The striking transition dynamics of breathers, new excitation
patterns for RWs, topological kink excitations, and other new stable excitation structures were obtained in the CNLS-p described systems.  We note that the CNLS-p system also admits cosine and sine wave solutions, which would provide a periodic wave background for nonlinear localized waves.  The periodic wave backgrounds constitute a stripe structure, which is similar to the ones obtained in a spin-orbital coupled spinor Bose-Einstein condensate \cite{Zhai}. This provides possibilities to investigate RW and other type nonlinear excitations on stripe phase background analytically and exactly. Here we focus on RW dynamics since the studies on RW in BEC system would be helpful for RW application and prevention in many other physical systems \cite{report}.

In this paper, we study RW excitation pattern on stripe phase background in a two-component BEC with pair-transition coupling effects. The dynamics of RW demonstrate some different behaviors compared with the previous ones in scalar and vector NLS described systems \cite{Akhmediev,cn,Chen,Zhao2,AKN,zhaoling1,zhaoling2}. Both RW pattern and RW numbers on temporal-spatial plane are much more abundant than the ones reported before. Particularly, we find the RW number  can be 1 to 16 except 5, 8, 14, and 15 numbers. Two RWs can be superimposed to admit the highest peak value which is nine times the background density, and six RWs admit the highest peak value which is 25 times the background density. But these value characters are admitted by the first-order RW (consist one RW) and the second-order RW (consist three RWs) of scalar NLS respectively. Therefore, the RW solutions here are not trivial superpositions of scalar NLS RW solutions.

Our presentations will be structured
as follows. In Sec. II,  we describe the theoretical model for a two-component BEC with pair-transition coupling effects and present stripe phase for the model which correspond to cosine and sine wave background in the two components.   In Sec. III, we demonstrate RW dynamics on the periodic wave background in the two components, based on derived rational solutions. In Sec. IV, the superposition cases for them are discussed, which demonstrate RW numbers on spatial-temporal distribution plane are $n (n+1)/2+m (m+1)/2$. Then in Sec. V, we discuss how to observe the RW dynamics here in a $^{87}Rb$ condensate with two hyperfine states. Finally,  we summarize the results and present our conclusions in Sec. VI.

\begin{figure}[htb]
\centering
\includegraphics[height=37mm,width=85mm]{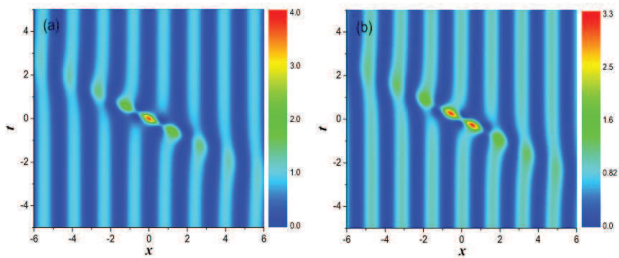}
\caption{(Color online) One rogue wave excitation on periodic wave background, (a) for component $q_1$ and (b) for component $q_2$. It is seen that rogue wave peaks are not nine times the background density value anymore.  The parameters are $a = 1, k = 2$. }\label{fig5}
\end{figure}

\begin{figure}[htb]
\centering
\includegraphics[height=37mm,width=85mm]{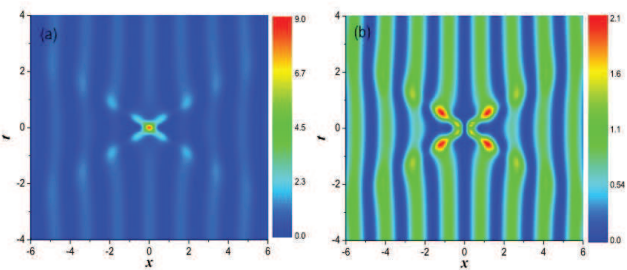}
\caption{(Color online) Two rogue waves excitation on periodic wave background, (a) for component $q_1$ and (b) for component $q_2$. It is seen that rogue wave peak is nine times the background density value in component $q_1$, but rouge wave peaks is much lower than nine times in component $q_2$.  The parameters are $a = 1, k = 2$. }\label{fig5}
\end{figure}

\section{The coupled Nonlinear Schr\"{o}dinger equations with particles' transition}

One-dimensional two-component BEC system with particle transition can be described by the Hamiltonian
$\hat{H}=\sum_j [-\frac{\hbar^2}{2m} \partial_x^2 \hat{q}_j \hat{q}_j^\dag+ \frac{g_{j,j}}{2}\hat{n}_j \hat{n}_j+ g_{j,3-j}\hat{n}_j \hat{n}_{3-j}
+J_1 (\hat{q}_j^\dag \hat{q}_{3-j}+\hat{q}_{3-j}^\dag \hat{q}_j)+\frac{J_2}{2} (\hat{q}_j^\dag \hat{q}_j^\dag \hat{q}_{3-j}\hat{q}_{3-j}+\hat{q}_{3-j}^\dag \hat{q}_{3-j}^\dag \hat{q}_j \hat{q}_j)]$ where $n_j=\hat{q}_j^\dag \hat{q}_j$ is the particle number operator, the symbol $^{\dag}$ represents the Hermite conjugation.
$g_{i,i} $ and $g_{3-i,i}$ $(i=1,2)$ are the intra and external interactions between atoms.
$J_1$ and $J_2$ denote single particle and pair particles transition coupling strength separately \cite{Fischer1,Fischer2}.
 In most studies,  $J_{1,2}$ are set to be zero usually because it was  believed that the presence of tunneling makes the systems become non-integrable \cite{Baronio,Zhao,Lakshmanan,Bludov}.
Recent experimental results in a double-well Bose-Einstein condensate suggested that pair-tunneling can become dominant with strong interaction between atoms  \cite{pair,Meyer}. Therefore, we
 consider that the case for second-order transition is dominant, namely, $J_1=0$ and $J_2\neq 0 $.  We find integrable CNLS-p can be derived from the Hamiltonian with $g_{j,3-j}=2 g_{j,j}=2 J_2$.

It is convenient to set $g_{j,j}=-\sigma$ ($\sigma=\pm 1$ correspond to attractive or repulsive interactions between atoms ) without losing generality for there is a trivial scalar transformation for different values.The corresponding dynamic evolution equation can be derived from the Heisenberg equation ${\rm i}\hbar(\partial \hat{q}_j/\partial t)=[\hat{q}_j,\hat{H}] $ for the field operator. Performing the mean field approximation $<\hat{q_j}>=q_j$, we can get the following integrable CNLS-p with scale dimensions $m=\hbar=1$
\begin{equation}\label{two-mode}
    \begin{split}
      {\rm i}q_{1,t}+\frac{1}{2}q_{1,xx}+(|q_1|^2+2|q_2|^2)q_1+ q_2^2\bar{q}_1&=0, \\
      {\rm i}q_{2,t}+\frac{1}{2}q_{2,xx}+ (2|q_1|^2+|q_2|^2)q_2+q_1^2\bar{q}_2&=0,
    \end{split}
\end{equation}
where the symbol overbar represents the complex conjugation. The coupled model can be also used to describe the propagation of orthogonally polarized
optical waves in an isotropic medium \cite{Boris}. Bond soliton fiber laser and soliton interactions were studied in a similar coupled model \cite{Tang}. The propagation of optical beams in terms of the two orthogonal modes of a planar waveguide where the beams
are only allowed to diffract in one spatial dimension can be described by the coupled model with some constrains on the ratio of cross- and self-phase modulation coefficients, cross-phase modulation and four-wave-mixing term  \cite{Menyuk,Kang}.  What needs mentioning is that the above coupled equations without the last term usually deem as non-integrable CNLS \cite{yang-benney}. However, when we add the particles transition term, the non-integrable CNLS become integrable, which was also proven by Painlev\'{e} analysis \cite{Park}.

\begin{figure}[htb]
\centering
\includegraphics[height=37mm,width=85mm]{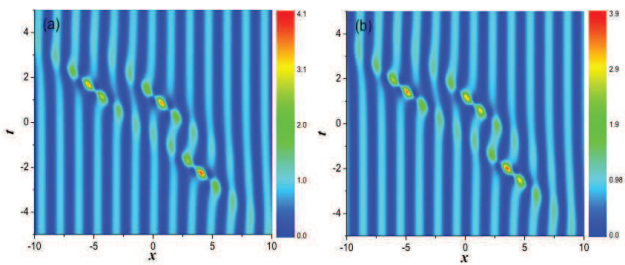}
\caption{(Color online) Three rogue waves excitation on periodic wave background, (a) for component $q_1$ and (b) for component $q_2$. The distribution profile is similar to the second-order rogue wave of scalar NLS reported before. The parameters are $a = 1, k = 2, c_1 = 5, c_2 = 10$. }\label{fig3}
\end{figure}

\begin{figure}[htb]
\centering
\includegraphics[height=37mm,width=85mm]{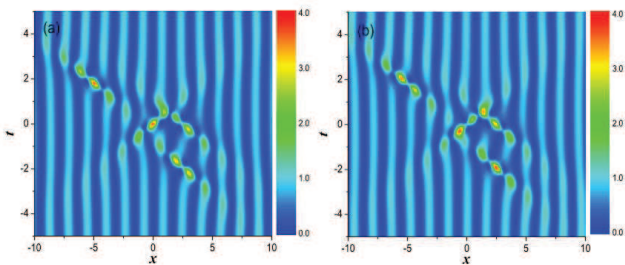}
\caption{(Color online) Four rogue waves excitation on periodic wave background, (a) for component $q_1$ and (b) for component $q_2$. The distribution profile is distinctive from the ones reported before. The parameters are  $a = 1, k = 2, c_1 = 10, c_2 = 0$. }\label{fig4}
\end{figure}

Recent studies on RWs suggest that
the unique wave arises from modulational instability, and the
rational solution of nonlinear partial equations
can be used to describe dynamics of RWs prototypically including both scalar and vector systems \cite{scalar1,scalar2,scalar3,spinor1,spinor2}. Among those different types of nonlinear partial equations, nonlinear Schr\"{o}dinger equation (NLS) has been given much attention because of its  widespread applications in optics, water wave tank, plasmas, and financial systems, as well as the quantum world of super fluids and Bose-Einstein condensate \cite{report}.
Based on the rational solutions, it has been found that there are
some different patterns for RWs, such as eye-shaped \cite{Akhmediev}, anti-eye-shaped \cite{Bludov,Zhao}, and four-petaled flower structure ones \cite{Zhao1}. The transition relation between them have been uncovered in coupled systems \cite{Chen,Zhao2}. The previous RW solutions are usually derived on plane wave backgrounds, but this does not mean that it can not exist on other types of  backgrounds. For example, some RW solutions on cnoidal wave backgrounds were presented in a semi-analytical forms \cite{cn}.  RW on different background would demonstrate some different dynamics properties \cite{Shin}. On the other hand, there are some certain laws for distribution numbers of RW on spatial-temporal distribution plane. The RW number is $n(n+1)/2$ for $n$-th-order RW solution of scalar NLS model \cite{AKN}, which means that there are
no higher-order solutions that are physically separable into
2,4,5,7,8,9,...elementary Peregrine RWs in scalar NLS described systems. Interestingly, we have shown that two or four RWs can exist in the coupled NLS described systems \cite{zhaoling1,zhaoling2}. Then, what about the RW pattern dynamics for the above CNLS-P model?

Bright soliton and RW were studied for the model that nonlinear coefficients and pair-transition coefficients are different \cite{Tian,BoTian}.
The RW excitation dynamics on a plane wave background have been discussed well \cite{zhao-ling,lingzhao}. But the general properties for RW number is still absent for the CNLS-P systems. We note that the CNLS-p admits periodic wave background. For example, $q_{10}= a\ cos(kx)\ e^{i \phi}$ and $q_{20}= -i a\ sin(kx)\ e^{i \phi}$ (where $\phi=a^2 t - k^2 t/2$ ) are also the solution of the CNLS-p model. The periodic wave backgrounds constitute a stripe structure, which is similar to the ones obtained in a spin-orbital coupled spinor Bose-Einstein condensate \cite{Zhai}. We here would like to study the excitation patterns of RW on these periodic wave backgrounds. We derive the RW solution exactly through performing the DT method \cite{lingzhao}. The two forms for DT can be used to construct single and double nonlinear localized waves separately. In the follows,  we list some non-trivial RW solutions for the system \eqref{two-mode}  according to RW numbers generated by the two DT forms.

\section{Different rogue wave excitation patterns on periodic wave backgrounds}

The solution for one RW is derived through the first DT form present in \cite{lingzhao} as
\begin{eqnarray}
q_{11}&=&  [\frac{2+4 i a^2 t}{4 a^4 t^2+4 a^2 (k t+x)^2+1}e^{-i k x}\nonumber\\
&&+i\ cos(kx-\pi/2) ]\ a\ e^{i \phi},\\
q_{21}&=& [\frac{2+4 i a^2 t}{4 a^4 t^2+4 a^2 (k t+x)^2+1}e^{-i k x}\nonumber\\
&&-sin(kx+\pi/2)]\ a\ e^{i \phi}.
\end{eqnarray}
We can see that the solution describe one RW on periodic wave background in the two components. For example, we show one case with $a=1$, $k=2$ in Fig. 1. It is seen that the peak value of RW about $4$ times the maximum value of background density in component $q_1$ and it is about $3.3$ times the maximum value of background density in component $q_2$. This is  different from the well-known RW in scalar NLS for which the peak of RW is nine times the background density value \cite{Akhmediev,AKN}. The superposition of them can be a RW on a plane wave background, and its value 5 times the background density value (see Fig. 5(a)) but it is still not nine times the background value. Therefore, the RW solutions here are not trivial superpositions of scalar RW solutions, which admit some different excitation dynamics.

The solution  for two RWs is derived through the second DT form present in \cite{lingzhao} as
\begin{eqnarray}
q_{12}&=&  [\frac{2+4 i a^2 t}{4 a^4 t^2+4 a^2 (k t+x)^2+1}e^{-i k x}\nonumber\\
&&+\frac{2+4 i a^2 t}{4 a^4 t^2+4 a^2 (-k t+x)^2+1}e^{i k x}\nonumber\\
&&- cos(kx) ]\ a\ e^{i \phi},\\
q_{22}&=& [\frac{2+4 i a^2 t}{4 a^4 t^2+4 a^2 (k t+x)^2+1}e^{-i k x}\nonumber\\
&&-\frac{2+4 i a^2 t}{4 a^4 t^2+4 a^2 (-k t+x)^2+1}e^{i k x}\nonumber\\
&&+i\ sin(kx)]\ a\ e^{i \phi}.
\end{eqnarray}
The solution describe two RWs on periodic wave background in the two components. For example, we show one case with $a = 1$, $k = 2$ in Fig. 2. It is seen that the peak value of the superposition of the two RWs is nine times the background density value in component $q_1$, but the peak value of RWs in component $q_2$ is much less than the nine times value. The RW structure is distinctive from the ones reported before, for which there are four humps around one center. This is also different from the four-petaled RW found in \cite{Zhao1}.

The solution for three RWs on periodic wave background is
\begin{eqnarray}
q_{13}&=& [-\frac{H(x,t)}{G(x,t)} e^{-i k x} + cos(kx)]\ a\ e^{i \phi}, \\
q_{23}&=& [-\frac{H(x,t)}{G(x,t)} e^{-i k x}  -\ i \ sin(kx)]\ a\ e^{i \phi},
\end{eqnarray}
where
\begin{widetext}
\small
\begin{eqnarray}
H(x,t)&=&6 i \left[32 a^{10} t^5+16 a^8 t^3 (4 k^2 t^2+t (8 k x-5 i)+4 x^2)+16 a^6 t \left(2 k^4 t^4+2 k^2 t^3 (4 k x-3 i)\right.\right.\nonumber\\
&&\left.\left.+t^2 (12 k^2 x^2-12 i k x+1)+2 t x^2 (4 k x-3 i)+2 x^4\right)+48 a^5 c_2 t^2\right.\nonumber\\
&&\left.-8 i a^4 \left(3 t^2 (4 i c_1 k+4 k^2 x^2-4 i k x+3)+2 t x (6 i c_1+x (4 k x-3 i))+2 k^4 t^4\right.\right.\nonumber\\
&&\left.\left.+2 k^2 t^3 (4 k x-3 i)+2 x^4\right)-48 a^3 c_2 \left(k^2 t^2+t (2 k x+i)+x^2\right)-6 i a^2 \left(t (8 c_1 k\right.\right.\nonumber\\
&&\left.\left.+8 k x-5 i)+4 x (2 c_1+x)+4 k^2 t^2\right)-12 a c_2+3 i\right],\nonumber\\
G(x,t)&=&64 a^{12} t^6+192 a^{10} t^4 (k t+x)^2+48 a^8 t^2 \left[4 k^4 t^4+16 k^3 t^3 x+3 t^2 (8 k^2 x^2+3)+16 k t x^3\right.\nonumber\\
&&\left.+4 x^4\right]+192 a^7 c_2 t^3+32 a^6 (k t+x) \left(18 c_1 t^2+2 k^5 t^5+10 k^4 t^4 x+20 k^3 t^3 x^2\right.\nonumber\\
&&\left.+20 k^2 t^2 x^3+k t (10 x^4-9 t^2)+x (2 x^4-9 t^2)\right)-576 a^5 c_2 t (k t+x)^2\nonumber\\&&+12 a^4 \left[-16 k^3 t^3 (c_1-x)+3 t^2 (-16 c_1 k^2 x+8 k^2 x^2+11)+16 k t x^2 (-3 c_1+x)\right.\nonumber\\&&\left.+4 x^3 (-4 c_1+x)+4 k^4 t^4\right]+432 a^3 c_2 t+36 a^2 [4 c_1^2+4 c_1 (k t+x)+4 c_2^2+3 (k t+x)^2]+9.\nonumber
 \end{eqnarray}
\end{widetext}
For an example, we show the dynamics of RWs with $a=1, k=2, c_1=5, c_2=10$ in Fig. 3. It is seen that there are three RWs on the spatial-temporal distribution plane. The distribution profile is identical with the second-order RW for scalar NLS, but the RW peak values are much smaller than the scalar ones.
\begin{figure*}[htb]
\centering
\includegraphics[height=70mm,width=165mm]{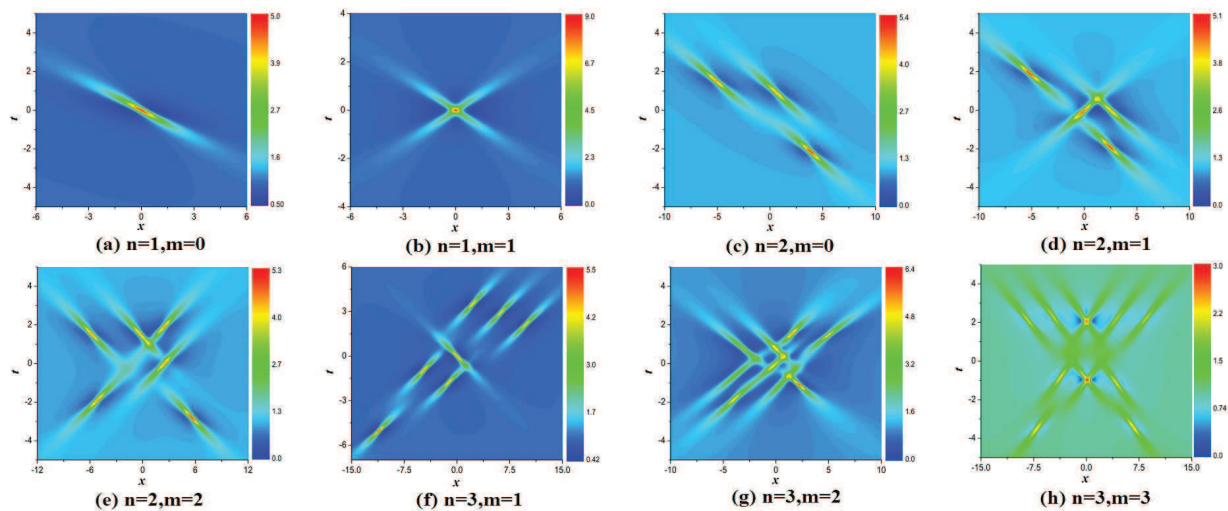}
\caption{(Color online) The dynamics evolution for the superposition of RWs in the two components. The rogue wave dynamics are given by the analytical form $|q_1|^2+|q_2|^2$ with different $n$ and $m$ values, where $q_1$ and $q_2$ can be written in combination forms of the $n$-th-order and $m$-th-order RW solutions of the scalar NLS. It is seen that the rogue waves distribution profiles are distinctive from the ones in scalar NLS described system and vector NLS descibed system with no pair-transition effects. }\label{fig5}
\end{figure*}

Similarly, we derive a rational solution which describes four RWs,
\begin{eqnarray}
q_{14}&=&[ - i \ sin(kx)-\frac{H(x,t)}{G(x,t)} e^{-i k x}\nonumber\\
&&+\frac{2+4 i a^2 t}{4 a^4 t^2+4 a^2 (-k t+x)^2+1}e^{i k x}]\ a\ e^{i \phi} ,\\
q_{24}&=&[ cos(kx)-\frac{H(x,t)}{G(x,t)} e^{-i k x}\nonumber\\
&&-\frac{2+4 i a^2 t}{4 a^4 t^2+4 a^2 (-k t+x)^2+1}e^{i k x}]\ a\ e^{i \phi}
\end{eqnarray}
which are the superposition of the first-order RW and the second-order RW. For an example, we show one case in Fig. 4. The profile of the four RWs can  be controlled by varying parameters $c_1$, $c_2$. The distribution profiles are much more abundant than the ones for coupled NLS without PT coupling effects \cite{zhaoling1,zhaoling2}.

More RWs numbers can be obtained by the iterations of DT. Particularly, we can further know that the RW number here can be $n (n+1)/2+m (m+1)/2$ (where $m,n $ are arbitrary non-negative integers), since the solutions are proven to be written in the form $q_1=\frac{\psi_{1n}+\psi_{2m}}{2}$ and $q_2=\frac{\psi_{1n}-\psi_{2m}}{2}$, where $\psi_{1n}$ and $\psi_{2m}$ are n-th-order and m-th-order RW solution on the $\psi_{10}=a\ e^{i [a^2 t-\frac{k^2 t}{2}-k x]} $ and $\psi_{20}=a\ e^{i [a^2 t-\frac{k^2 t}{2}+k x]}$ backgrounds respectively for the scalar NLS ${\rm i}\ \psi_{j,t}+\frac{1}{2}\psi_{j,xx}+|\psi_j|^2 \psi_j =0 $ ($j=1,2$). The RW number is $n(n+1)/2$ for $n$-th-order RW of scalar NLS system \cite{AKN}. As shown in \cite{Park}, the Eq. (1) can be related with two uncoupled NLS equations through a simple linear transformation, which can be also used to construct the solutions here from the ones of scalar NLS. This explains why the linear superposition
of two RWs on different plane background is possible in above results. We have presented them with the first-order and second-order in the above discussions.  Higher-order RW solution can be derived similarly \cite{lingzhao,guoling}, we do not show them here because of their much complicated expressions.

\section{Different rogue wave excitation numbers on a superposition plane wave background}
The superposition of the rogue wave in the two components describe RWs on plane wave backgrounds, since the cosine and sine wave backgrounds are summed to be a plane wave. This is similar to the RWs in scalar NLS described systems. However, there are still many differences between them. Firstly, the highest peak is lower. Secondly, the RW numbers are quite different. The number of RW can be $n (n+1)/2+m (m+1)/2$ (where $m, n$ are non-negative integers), in contrast to $n(n+1)/2$ for scalar NLS systems. For example, we summarize them to the fourth-order in the table I. We can see that RW number can be 0 to 16 except $5,8,14$ and $15$. These characters are distinctive from the ones in scalar NLS and even two-component NLS without transition effects.
\begin{table}[!h]
\centering
\small
 \label{table1}
\begin{tabular}{|c|c|}
 \hline
  Existence condition & \ \  Rogue wave number \ \ \\
  \hline
  $n=0,m=0$ & 0 \\
\hline
  $n=1,m=0$ & 1  \\
\hline
 $n=1,m=1$ & 2 \\
  \hline
  $n=2,m=0$ & 3  \\
  \hline
  $n=2,m=1$ & 4 \\
  \hline
 $n=2,m=2$ or $n=3,m=0$
  & 6\\
  \hline
   $n=3,m=1$ & 7  \\
  \hline
  $n=3,m=2$ & 9 \\
  \hline
  $n=4,m=0$ & 10  \\
  \hline
  $n=4,m=1$ & 11 \\
  \hline
  $n=3,m=3$ & 12 \\
  \hline
  $n=4,m=2$ & 13  \\
  \hline
  $n=4,m=3$ & 16  \\
  \hline
\end{tabular}
\caption{The possible numbers for fundamental rogue waves emerging on the spatial-temporal distribution plane up to $n=4$ and $m=3$ case. It is seen that the numbers $5, 8, 14$, and $15$ are absent in the pair-transition coupled model. }
\end{table}
\begin{figure}[htb]
\centering
\label{fig:7}
{\includegraphics[height=37mm,width=85mm]{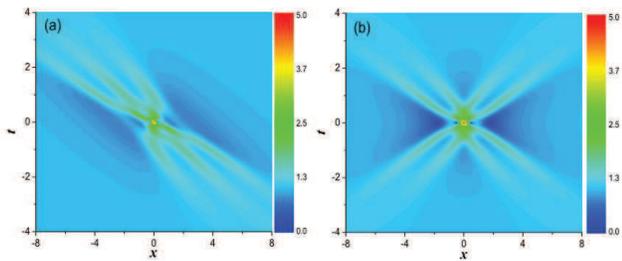}}
\caption{(color online) (a) The superposition $\sqrt{|q_1|^2+|q_2|^2}$ of six rogue waves with $n=3, m=0$ which admits the highest peak value. (b) The superposition $\sqrt{|q_1|^2+|q_2|^2}$ of six rogue waves with $n=2, m=2$ which admits the highest peak value. It is seen that the highest peaks in the two cases are both 25 times the background density value, but the distribution profile with $n=3, m=0$ is distinctive from the case with $n=2,m=2$.}
\end{figure}
\begin{figure}[htb]
\centering
\label{fig:8}
{\includegraphics[height=37mm,width=85mm]{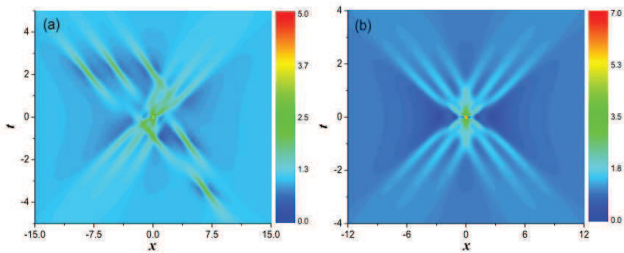}}
\caption{(color online)  Two different cases for twelve rogue waves with $n=3, m=3$ shown in $\sqrt{|q_1|^2+|q_2|^2}$. (a) Six separated rogue waves and the superposition of six rogue waves with the highest peak. (b) The superposition of twelve rogue waves with the highest peak, which is 49 times the background density value. }
\end{figure}

We show the dynamics of them in Fig. 5, based on the RW solutions to the third-order case. It is seen that the distribution patterns are distinctive from the ones of scalar NLS in \cite{AKN,guoling,He} which admit certain structures, such as structural translations, triangular cascades, pentagrams, heptagrams, enneagrams, etc.. This comes from that there are more free parameters here which brings more abundant structures.
Particularly, the same number can be obtained by different excitation forms. For example, the number six can be obtained by $n=2, m=2$, but it can be also obtained by $n=3, m=0$. The distribution characters between them are distinctive (see Fig. 6 for the superposition case with the highest peak value), because the case $n=2, m=2$ has more free parameters than the case $n=3, m=0$. The case $n=3, m=0$ admits identical patterns with the third-order RW of scalar NLS. But the RW peak values are different between them.

The locations of RWs can be changed, and superposed to admit different peak values. For two RWs case, the highest peak can be nine times the background density (see Fig. 5(b)), which is identical with the first-order RW for scalar NLS system. But this is superposed by two RWs, and the first-order RW for scalar NLS has one RW. For six RWs case, the highest peak of RWs can be 25 times background density (see Fig. 6 (a) and (b)), which is identical with the second-order RW for scalar NLS system. The second-order scalar RW admit three RWs, in contrast to six RWs here. The superposition form can be also  much more abundant than the cases for scalar system. For example, we show two different cases for the twelve RWs with varying parameters in Fig. 7. In Fig. 7(a), there are six separate RWs and superposition of six RWs with the highest peak. In Fig. 7(b), there is a superposition of twelve RWs with the highest peak. It is shown that the highest peak of the twelve RWs superposition is 49 times the background density value, which is identical with the six RWs superposition case of the third-order RW for scalar NLS.  All these characters indicate that the RW solutions here are not trivial superpositions of scalar RW solutions.

\begin{figure}[htb]
\centering
\label{fig:8}
{\includegraphics[height=37mm,width=85mm]{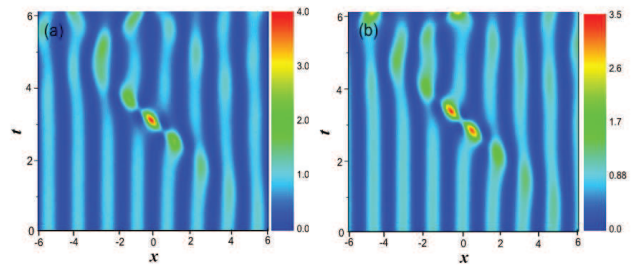}}
\caption{(color online): The numerical simulation of a rogue wave on periodic wave background with small noise. The initial excitation condition is given by the exact ones at $t=-3$ in Fig. 1 by multiplying a factor $(1+0.01 Random[-1,1])$. It is seen that the rogue waves are robust against small noise. }\label{fig8}
\end{figure}

\section{Possibilities to observe these nonlinear excitations}
The pair-transition(PT) term corresponds to pair particles transition in two-component Bose-Einstein condensate \cite{Fischer1,Fischer2} or four-wave-mixing effect in a nonlinear planar waveguide \cite{Menyuk,Kang}.  Therefore, they can be realized in a two-component ultra-cold atomic system or planar waveguide with two orthogonal modes through combining these intensity and phase modulation techniques. As an example, we discuss possibilities to observe the rogue wave in Fig. 1 in a cigar-shaped condensate with two hyperfine
states, $q_1$ and $q_2$.

For simplicity, we assume the initial
condensation occurring in the trapped state $q_2$. State $q_1$
is coupled to $q_2$ by an RF or microwave field tuned near the
$q_2 \rightarrow q_1$ transition. The PT effects can be
realized by the RF field in the strong interaction regimes
 \cite{pair,Meyer}. The total number of $\,^{87}Rb$ atoms in
the condensate is $N= 5\times 10^4$. $a_{i,j}$ $(i,j=1,2)$ are s-wave
scattering lengths which can be adjusted by Feshbach resonance
technique. Setting $a_{1,2}=a_{2,1}=1.6\ nm$ and
$a_{2,2}=a_{1,1}=0.8 \ nm$, under mean-field approximation, the
s-wave scattering effective interaction strengths between atoms in
the same hyperfine state are $U_{j,j}=4 \pi \hbar^2 a_{j,j} /m$ ($m$
is the atom mass), and the scattering effective interaction
strengths between atoms in different hyperfine state are
$U_{j,3-j}=4 \pi \hbar^2 a_{j,3-j} /m$.
 When the interaction between atoms is
attractive and the PT coefficient is $N\cdot U_{1,1}$, the units in
axial direction and time are scaled to be $2.0\ \mu m$ and $0.5 \
ms$ respectively, the dynamics of the condensate with PT effects can
be described well by the Eq. (1). Recently, rogue wave and Akhmediev breather have been excited experimentally in nonlinear fiber system under the direction of related exact solutions \cite{RW,Kibler}.  Vector soliton including dark-dark, bright-dark soliton, bright-bright soliton, and even half-soliton have been excited experimentally in multi-component Bose-Einstein condensate based on density and phase modulation techniques \cite{Engels}. The experiments indicated that the initial conditions for these nonlinear excitations can be made nearly precisely by density and phase modulation techniques. Similarly, the exact solution (2) for rogue wave on sine wave background can be used to direct initial density and phase modulation explicitly in the
two components.

However, the initial condition can not be made precisely. There are always some deviations in real experiments. Therefore, we test the evolution of these nonlinear excitations with some noises. The results indicate that they are robust against small noises or perturbations. For an example, we show the results for one RW on sine and cosine wave background in Fig. 8.  The time of transition process is about $1.0 \ ms$ for the rogue wave. The time duration is much shorter than the life time of a Bose condensate. Therefore, the RWs can be observed from the initial conditions approaching to the ideal ones given by these exact solutions in the two-component condensate system.

\section{Conclusion and discussion}
In conclusion,  the RWs on stipe phase background are reported in a two-component BEC with pair-transition coupling effects, which demonstrate some different behaviors compared with the previous ones in scalar and vector NLS described systems. Both RW pattern and RW numbers on temporal-spatial plane are much more abundant than the ones reported before. Particularly, we find the RW number are $n (n+1)/2+m (m+1)/2$ (where $m, n $ are arbitrary non-negative integers), which generates 1 to 16 except 5, 8, 14, and 15 numbers. These characters and the RW peak values suggest that the RW solutions here are not trivial superpositions of scalar NLS RW solutions. It should be noted that the cosine wave background is distinctive from the fluctuating periodic wave background studied in \cite{Shin}, which is a weak fluctuation on a plane wave background.  Especially, the RWs solution presented here can be obtained from the ones of standard NLS on plane wave background with different wave vectors, since the CNLS-p model can be transformed to two uncoupled NLS equations through a linear transformation \cite{Park}. The solutions of Eq. (1) can be written in the form $q_1=\frac{\psi_{1n}+\psi_{2m}}{2}$ and $q_2=\frac{\psi_{1n}-\psi_{2m}}{2}$, where $\psi_{1n}$ and $\psi_{2m}$ are n-th-order and m-th-order RW solution for the scalar NLS ${\rm i}\ \psi_{j,t}+\frac{1}{2}\psi_{j,xx}+|\psi_j|^2 \psi_j =0 $ ($j=1,2$). For an example, the solution (Eq. (2) and (3)) can be obtained from a fundamental RW solution $\psi_{11}$ on a plane wave background ($\psi_{10}=a\ e^{i [a^2 t-\frac{k^2 t}{2}-k x]} $) and a plane wave solution $\psi_{20}=a\ e^{i [a^2 t-\frac{k^2 t}{2}+k x]}$ for the scalar NLS. Then $q_{11}=\frac{\psi_{11}+\psi_{20}}{2}$ and $q_{21}=\frac{\psi_{11}-\psi_{20}}{2}$ can be simplified directly to be the solution Eq. (2) and (3). Explicitly, $\frac{\psi_{10}\pm \psi_{20}}{2}$ induce cosine or sine wave background and the RW signal comes from the RW signal in $\psi_{11}$ ($\psi_{11}$ can be written as $\psi_{signal}+\psi_{10}$ form). Therefore the simple transformation is nontrivial, which can be used to construct these different excitation patterns for vector RW, and other types localized waves. The results here create  opportunities to study transition dynamics of nonlinear localized waves on periodic wave background exactly and analytically.

\section*{Acknowledgments}
This work is supported by National Natural Science Foundation of
China (Contact No. 11405129, 11401221), and Shaanxi Province Science association of colleges and universities (Contact No. 20160216).


\begin{thebibliography}{99}
\bibitem{Kevrekidis} P. G. Kevrekidis, D. Frantzeskakis, and R. Carretero-
Gonzalez, Emergent Nonlinear Phenomena in Bose-Einstein
Condensates: Theory and Experiment (Springer, Berlin Heidelberg,
2009).
\bibitem{Wu} B. Wu, J. Liu,  and Q. Niu
, ``Controlled generation of dark solitons with phase imprinting", Phys. Rev. Lett. 88, 034101 (2002).
\bibitem{Matuszewski} M. Matuszewski, E. Infeld, B.A. Malomed, et al., ``Fully three dimensional breather solitons can be created using Feshbach resonances", Phys.
Rev. Lett. 95, 050403 (2005).
\bibitem{Bludov1} Yu. V. Bludov, V. V. Konotop, and N. Akhmediev, Phys. Rev. A 80, 033610 (2009).
\bibitem{Mrwb} K. Manikandan, P. Muruganandam, M. Senthilvelan, and M. Lakshmanan
Phys. Rev. E 90, 062905 (2014).
\bibitem{spinbec} Y. Kawaguchi, M. Ueda, Phys. Rep. 520, 253 (2012).
 \bibitem{Becker} C. Becker, S. Stellmer, P.S. Panahi, S. Dorscher, M. Baumert,
Eva-Maria Richter, J. Kronjager, K. Bongs, K. Sengstock, Nature
Phys. 4,  496-501 (2008).
\bibitem{Engels}  C. Hamner, J. J. Chang, and P. Engels,
Phys. Rev. Lett. 106, 065302 (2011); M. A. Hoefer, J. J. Chang, C.
Hamner, and P. Engels, Phys. Rev. A 84, 041605(R) (2011).
\bibitem{Bludov} Y.V. Bludov, V.V. Konotop, and N. Akhmediev, Eur. Phys.
J. Special Topics 185, 169 (2010).
\bibitem{Zhao} L.C. Zhao, J. Liu, J. Opt. Soc. Am.
B 29,  3119-3127 (2012).
\bibitem{Zhao1} L.C. Zhao, J. Liu, Phys. Rev. E 87, 013201
(2013).
\bibitem{Baronio} F. Baronio, A. Degasperis, M. Conforti, and S. Wabnitz, Phys. Rev. Lett. 109,  044102 (2012).
\bibitem{Zhao2} L.C. Zhao, G.G. Xin, Z.Y. Yang, Phys. Rev. E 90, 022918 (2014).
\bibitem{Lakshmanan} T. Kanna and M. Lakshmanan, Phys. Rev. Lett. 86,  5043-5046(2001)
; M. Vijayajayanthi, T. Kanna, and M. Lakshmanan, Phys. Rev. A
77,  013820 (2008).
\bibitem{Guo} B.L. Guo and L.M. Ling, Chin. Phys. Lett. 28, 110202
(2011).
\bibitem{Liu} J. Liu, L. Fu, B.Y. Ou, S.G. Chen, et al., Phys. Rev. A 66, 023404 (2002).
\bibitem{Wadati} J. Ieda, T. Miyakawa, and M. Wadati, Phys. Rev. Lett. 93,
194102  (2004).
\bibitem{Qin} Z.J. Qin, G. Mu, Phys. Rev. E 86, 036601 (2012).
\bibitem{Lahini} Y. Lahini, F. Pozzi, M. Sorel, R. Morandotti, D. N. Christodoulides,
and Y. Silberberg, Phys. Rev. Lett. 101,  193901 (2008).
\bibitem{Li} Y.Y. Li, W. Pang, S.H. Fu, and B. A. Malomed, Phys. Rev. A 85,  053821 (2012).
\bibitem{Williams} J. Williams, R. Walser, J. Cooper, et al., Phys.
Rev. A 59,  R31 (1999); J. Williams, R. Walser, J. Cooper, E. A.
Cornell, and M. Holland, Phys. Rev. A 61, 033612 (2000).
\bibitem{Fischer} U.R. Fischer, C. Iniotakis, and A. Posazhennikova, Phys. Rev. A 77, 031602(R) (2008).
\bibitem{Fischer1} P. Bader, U.R. Fischer, Phys. Rev. Lett. 103, 060402 (2009).
\bibitem{Fischer2} U.R. Fischer, K.S. Lee, B. Xiong, Phys. Rev. A 84, 011604 (2011).
\bibitem{pair} S. F$\ddot{o}$lling, S.
Trotzky, P. Cheinet, et al., Nature 448,  06112 (2007).
\bibitem{Meyer} S. Z\"{o}llner, H.D. Meyer, and P. Schmelcher, Phys. Rev. Lett.
100,  040401 (2008).
\bibitem{zhao-ling} L.C. Zhao, L. Ling, Z.Y. Yang, J. Liu, Commun Nonlinear Sci Numer Simulat 23,  21-27 (2015).
\bibitem{Park} Q-Han Park and H.J. Shin, Phys. Rev. E 59, 2373 (1999).
\bibitem{Tian} X. L\"{u}, and B.  Tian,  Phys. Rev. E 85, 026117 (2012).
\bibitem{BoTian} W.R. Sun, B. Tian, Y. Jiang, and H.L. Zhen, Phys. Rev. E 91, 023205 (2015).
\bibitem{lingzhao} L. Ling and L.C. Zhao, Phys. Rev. E 92, 022924 (2015).
\bibitem{Zhai} C. Wang, C. Gao, C.M. Jian, H. Zhai, Phys. Rev. Lett. 105, 160403 (2010).
\bibitem{report} M. Onorato, S. Residori, U. Bortolozzo, A. Montina, and F. T.
Arecchi, Phys. Rep. 528, 47 (2013).
\bibitem{Chen} S.H. Chen, P. Grelu, and J. M. Soto-Crespo,Phys. Rev. E 89,
011201(R) (2014).
\bibitem{AKN} D. J. Kedziora, A. Ankiewicz, and N. Akhmediev, Phys. Rev. E 88, 013207 (2013).
\bibitem{zhaoling1} L. Ling, B. Guo, and L.C. Zhao, Phys. Rev. E 89, 041201(R) (2014).
\bibitem{zhaoling2} L.C. Zhao, B. Guo, and L. Ling, J. Math. Phys. 57, 043508 (2016).
\bibitem{Akhmediev} N. Akhmediev, A. Ankiewicz, and J. M. Soto-Crespo, Phys.
Rev. E 80, 026601(2009).

\bibitem{cn} D.J. Kedziora, A. Ankiewicz, and N. Akhmediev, Eur. Phys. J. Special Topics 223, 43–62 (2014).

\bibitem{Boris} Boris A. Malomed, Phys. Rev. A 45, R8321 (1992).
\bibitem{Tang} D.Y. Tang, B. Zhao, D.Y. Shen, et al.,
Phys. Rev. A 66, 033806 (2002).

\bibitem{Menyuk} C. R. Menyuk, IEEE J. Quantum Electron. QE-23, 174
(1987).
\bibitem{Kang} J.U. Kang and G.I. Stegeman, J.S. Aitchison, N. Akhmediev, Phys. Rev. Lett. 76,
3699 (1996).
\bibitem{yang-benney}  J. Yang and D. J. Benney, Stud. Appl. Math. 96, 111 (1996).

 \bibitem{scalar1} B. Kibler, J. Fatome, C. Finot, et al., Nat. Phys. 6, 790–795 (2010).
\bibitem{scalar2} A. Chabchoub and N. Akhmediev,  Phys. Lett. A 377, 2590–2593 (2013).
\bibitem{scalar3} A. Chabchoub, N. Hoffmann, M. Onorato, and N. Akhmediev,
Phys. Rev. X 2, 011015 (2012).
\bibitem{spinor1} F. Baronio, M. Conforti, A. Degasperis, et al., Phys. Rev. Lett. 113, 034101 (2014).
\bibitem{spinor2} B. Frisquet, B. Kibler, P. Morin,  et al., Sci Rep. 6, 20785 (2016).
\bibitem{Shin} H. J. Shin, Phys. Rev. E 88, 032919 (2013).
\bibitem{guoling} B.L. Guo, L.M. Ling, Q.P. Liu, Phys. Rev. E 85, 026607 (2012).
\bibitem{He} J.S. He, H.R. Zhang, L.H. Wang, K. Porsezian,
and A.S. Fokas, Phys. Rev. E 87, 052914 (2013).
\bibitem{RW} J. M. Dudley, G. Genty, F. Dias, B. Kibler, and N. Akhmediev, Opt. Express 17, 21497 (2009) .
\bibitem{Kibler}  B. Kibler, J. Fatome, C. Finot, G. Millot, et al., Sci. Rep. 2, 463 (2012).
\end{thebibliography}
\end{document}